\documentclass[a4paper]{llncs}

\usepackage[utf8]{inputenc}
\usepackage[T1]{fontenc}

\usepackage{amssymb,amsmath}
\usepackage{float}
\usepackage{graphicx,color,layout}
\usepackage{enumerate}
\usepackage{xspace}
\usepackage{multicol}
\usepackage{subcaption}

\newcommand{\ignore}[1]{{}}

\newcommand{\be}{\begin{equation}}
\newcommand{\ee}{\end{equation}}
\newcommand{\beq}{\begin{equation}}
\newcommand{\eeq}{\end{equation}}
\newcommand{\bea}{\begin{eqnarray}\displaystyle}
\newcommand{\eea}{\end{eqnarray}}

\usepackage[ruled, vlined]{algorithm2e}
\SetKwInput{KwInput}{Input} 
\SetKwInput{KwOutput}{Output} 
\SetKw{KwBy}{by}
\SetKw{KwTo}{to}
\usepackage{adjustbox}

\newcommand{\ie}{i.e.\xspace}

\title{On-the-fly Optimization of Parallel Computation of Symbolic Symplectic Invariants}

\author{Joseph Ben Geloun\inst{1}, Camille Coti\inst{1}, Allen D. Malony\inst{2}}
\institute{%
LIPN, CNRS UMR 7030, Universit\'e Sorbonne Paris Nord \\
99, avenue Jean-Baptiste Cl\'ement \\ F-93430 Villetaneuse, FRANCE\\
\email{\{joseph.bengeloun,camille.coti\}@lipn.univ-paris13.fr}
}
\institute{%
LIPN, CNRS UMR 7030, Universit\'e Sorbonne Paris Nord \\
99, avenue Jean-Baptiste Cl\'ement \\ F-93430 Villetaneuse, FRANCE\\
\email{\{joseph.bengeloun,camille.coti\}@lipn.univ-paris13.fr}
  \and University of Oregon\\
  \email{malony@cs.uoregon.edu}
}

\begin{document}

\maketitle

\begin{abstract}
Group invariants are used in high energy physics to define quantum field theory interactions. In this paper, we are presenting the parallel algebraic computation of special invariants called symplectic and even focusing on one particular invariant that finds recent interest in physics. Our results will export to other invariants. The cost of performing basic computations on the multivariate polynomials involved evolves during the computation, as the polynomials get larger or with an increasing number of terms. However, in some cases, they stay small. Traditionally, high-performance software is optimized by running it on a smaller data set in order to use profiling information to set some tuning parameters. Since the (communication and computation) costs evolve during the computation, the first iterations of the computation might not be representative of the rest of the computation and this approach cannot be applied in this case. To cope with this evolution, we are presenting an approach to get performance data and tune the algorithm during the execution.
\end{abstract}

\section{Introduction}

Many scientific applications that run in parallel on high-performance computing (HPC) systems display a regular execution behavior whose performance is relatively stable during the course of the computation, thereby making it possible to empirically analyze and model for post-mortem tuning purposes.  While performance can certainly depend on parallelism degree, problem size, and other factors, if the algorithms used generate computational work that is predictable, repeatable, and does not vary in unknown manners during execution, there is a good chance that static optimizations and judicious configuration settings will be effective.

Now suppose that this is not the case.  Consider an application whose  computational behavior is more irregular and even non-deterministic, whose state defines the amount and type of work to be done next, where execution dependencies change as a result, and when termination is subject to varying conditions that are difficult to assess.  Now imagine that the problem size is also a factor in how the application unfolds.  Such an application can cause havoc in relation to performance observation, analysis, and improvement.  More to the point, it raises serious concern whether a post-mortem methodology even applies.

In such a scenario, it might be fair to argue that the performance problem becomes less a matter static code optimization and more an issue of dynamic management of resources and parallelism at runtime.  In the presence of changing execution behavior and varying computational demands, it becomes more necessary to address concurrency and resource interactions dynamically in order to control inefficiences that manifest themselves at runtime.  If the performance is not predictable from run to run, post-mortem empirical analysis will prove even further inadequate.

In this paper, we look at group invariants and their use in high energy physics to define quantum field theory interactions. In particular, we  present the parallel algebraic computation of special invariants called \textit{symplectic} and focus on one particular invariant that finds recent interest in physics.  In successful, our results will export to other invariants.  Interestingly, the cost of performing basic computations on the multivariate polynomials involved evolves during the computation, as the polynomials get larger or with an increasing number of terms.  However, in some cases, they stay small.  Since the (communication and computation) costs change during the computation, the first iterations of the computation might not be representative of the rest of the computation.  Hence, traditional techniques of tuning performance based on prior profiling runs with smaller data sets is unlikely to prove fruitful.

 To cope with this evolution, we are presenting an approach to observe performance online and make adaptive optimization decisions on-the-fly during the execution.  In particular, we focus on the control of parallelism degree as informed by monitoring the concurrency availability, execution overheads, and workload balance.  When we identify execution states offering opportunities for performance improvement, adaptive control policies are engaged and evaluated.
 
 Our contributions are found both in the parallel algorithm and implementation of symbolic symplectic invariants computations, as well as the on-the-fly optimization methodology and its development.  We show results that demonstrate the ability to scale the problem, delivering outcomes heretofore unachieved, while maintaining good speedup returns.
 The paper begins with a more formal description of the problem.  We then discuss the parallelization approach and the runtime performance observaton and control.  Experimental benchmarks and performance results are presented, with the paper concluding with summary observations and future directions.

 \section{Simplectic Invariants}
\label{sec:invar}

\subsection{Group invariants and tensor models}
\label{sec:invar:physics}

In high energy physics, one
ordinarily uses \textit{group invariants} to define quantum field theory interactions. 
More recently, particularly among the attempts of quantizing gravity, \textit{tensor models} \cite{ambj3dqg,Guraubook} have extensively addressed classical Lie group (unitary and orthogonal) invariants in their construction. 
Tensor models belong to an active domain 
that stems from the proposal of making random and discrete the space or spacetime and the study of their continuum limit. 
They naturally generalize, in higher dimension, the so-called \textit{matrix models} \cite{DiFrancesco:1993nw} focusing on gravity in 2D.

The physics behind tensor models strongly rests on their rich combinatorics.  
Indeed, the interactions of tensor models, that are tensor contractions, correspond
to classical Lie groups invariants and encode in terms of edge-colored graphs.  With complex tensors,  
one deals with  $U(N)^{\otimes d}$ (unitary) invariants \cite{Bonzom:2012hw}. 
Meanwhile, for the real tensors, these interactions map to $O(N)^{\otimes d}$ (orthogonal) invariants \cite{Carrozza:2015adg}. 
The enumeration of these Lie group invariants was addressed in a series of works \cite{Sanjo,BenGeloun:2017vwn,Avohou:2019qrl}.  
As upshots of these analyses, the enumeration rules of tensor invariants shows new correspondences with branched covers in algebraic geometry and topology, 
and, using representation theoretic formulae, the same counting expresses
in terms of the famous Kronecker coefficient, an object of paramount importance 
in computational complexity theory \cite{stan,iken1,Blasiak}. 
 
Alongside with the orthogonal and unitary groups, one of the most prominent classical Lie groups is certainly the \textit{symplectic group} $Sp(2N)$. 
Given the aforementioned results, a natural question follows: what are the enumeration rules satisfied by the symplectic invariants?
This question is far from being purely formal. 
Indeed, some recent problems in condensed matter/black hole physics
 \cite{Carrozza:2018psc} show that the symplectic group plays therein an important role. 
Thus, the systematic study of the $Sp(2N)$ invariants becomes urgent for physics.  
 
The calculation of symplectic invariants, 
as opposed to unitary and orthogonal invariants, 
contains sign factors due to the symplectic matrix $J$ that they involve. 
Therefore, nothing prevents that a given 
tensor contraction yields an identically vanishing invariant.  It is therefore pertinent to test
at small order in $N$, and small order in 
the number of tensor contracted, if a symplectic invariants is simply not zero.  Something 
important that one must bear in mind is
that the invariant is a polynomial 
of formal (real) variables made out of the tensor components. 
Any program aiming at evaluating these invariants should
perform formal calculus. 

At rank $d=3$, one contraction 
of 4 tensors is particularly interesting for physicists: the so-called \textit{complete graph contraction} (for instance see \cite{Carrozza:2015adg}). 
This invariant may not exist for the symplectic
group.  One of the purposes of the present work 
is to show that up to order $2N=8$, this invariant
is not identically null. 

\subsection{Basics on polynomial symplectic invariants}
\label{sec:invar:basic}

We set up now our notation. 
The definition of polynomial symplectic invariants requires
the real $2N\times 2N$ symplectic matrix $J$ which expresses in blocks
\bea
 J  =\left(\begin{array}{cc}
0 & I_N \cr
- I_N & 0
\end{array}
\right)\,, 
\qquad J^2 = - I_{2N} \,, 
\eea
where $I_N$, for all $N$, is the identity matrix of $M_N(\mathbb{R})$. 
A matrix $K \in Sp(2N)$ obeys $K J K^{T} = J, $ and  $K^T J K = J$. 

A rank-$d$ real tensor $T$, with components  $T_{p_1,\dots, p_d}$, $p_c=1,\dots, 2N_c$, transforms under the fundamental 
representation of $\otimes_{c=1}^d Sp(2N_c)$, if for fixed $N_c$, if the group $Sp(2N_c)$ 
acts on the index $p_c$ such that the transformed tensor satisfies: 
\be
T^{K}_{q_1,\dots, q_d} = 
\sum_{p_1,\dots p_d} 
K^{(1)}_{q_1p_1} \dots K^{(d)}_{q_1p_1} \;
T_{p_1,\dots  , p_d} \,, 
\ee
where $K^{(c)} \in Sp(2N_c)$, $c=1,\dots, d$. 
For simplicity, in the rest of this work, we assume
$N_c = N$, for all copies.

The interactions of $Sp(2N)$ tensor models consists in the contractions of an even number of tensors $T$ 
where the contraction metric  is
precisely the matrix $J$.  It is not complicated to show that
they are invariant under the fundamental representation of $Sp(2N)^{\otimes d}$.

Among these invariants, 
there is a particular physical interest in the following
contraction of four tensors: 
\bea\label{T4form}
T^4 &= & \sum_{a_1,a_2, \dots, a_6}
\sum_{\bar a_1, \bar a_2, \dots, \bar a_6}
J_{a_1\bar a_1}J_{a_2\bar a_2}
J_{a_3\bar a_3}J_{a_4\bar a_4}
J_{a_5\bar a_5}J_{a_6\bar a_6}  \crcr
&\times &
T_{a_1,a_2,a_3}T_{a_4,a_5,\bar a_3}
T_{\bar a_4,\bar a_2,a_6}T_{\bar a_1, \bar a_5, \bar a_6} 
\eea

with, for all $i$, $a_i$ and $\bar a_i$ ranging from 1 to $2N$. 
On the left-hand side of the equality, $T^4$
is mere notation for that invariant, that can be
represented by the complete graph. 

We will see the most naive way of implementing the evaluation of tensor contractions in section \ref{sec:algo:naive}.  However, with 12 nested loops, it is computationally complex and time expensive.  Hence, we need to extract algebraic properties that reduce this complexity.

Using the properties of the matrix $J$, the above sum 
reduces to 
\bea
&&
 T^4 = 
 \prod_{l=1}^6
 \Big[\sum_{a_l=1}^{N}\sum_{\bar a_l=N+1}^{2N}
 J_{a_ l \bar a_l } 
 + 
 \sum_{a_l=N+1}^{2N}\sum_{\bar a_l=1}^{N}
 J_{a_ l \bar a_l }
 \Big]
T_{a_1,a_2,a_3}T_{a_4,a_5,\bar a_3}
T_{\bar a_4,\bar a_2,a_6}T_{\bar a_1, \bar a_4, \bar a_6}
\crcr
&&
=   \sum_{I \subset \{1,2,\dots, 6\}}
\prod_{l\in I}
 \Big[\sum_{a_l=1}^{N}\sum_{\bar a_l=N+1}^{2N}
 J_{a_ l \bar a_l }  \Big]
\prod_{l\notin I}
 \Big[
 \sum_{a_l=N+1}^{2N}\sum_{\bar a_l=1}^{N}
 J_{a_ l \bar a_l }
 \Big] \crcr
 && \qquad \times 
T_{a_1,a_2,a_3}T_{a_4,a_5,\bar a_3}
T_{\bar a_4,\bar a_2,a_6}T_{\bar a_1, \bar a_4, \bar a_6}
\cr\cr
&&
= \sum_{I \subset \{1,2,\dots, 6\}}
(-1)^{6-|I|}
\prod_{l\in I}
 \Big[\sum_{a_l=1}^{N}\sum_{\bar a_l=1}^{N}
\delta_{\bar a_l, a_l + N}  \Big]
\prod_{l\notin I}
 \Big[
 \sum_{a_l=1}^{N}\sum_{\bar a_l=1}^{N}
\delta_{a_l, \bar a_l + N}
 \Big] \crcr
&& \qquad \times
T_{a_1,a_2,a_3}T_{a_4,a_5,\bar a_3}
T_{\bar a_4,\bar a_2,a_6}T_{\bar a_1, \bar a_4, \bar a_6}\,.
\label{sumI}
\eea
That sum breaks in $2^{6}$ sub-sums. 
Fortunately, we use the symmetry of the pattern 
$T_{a_1,a_2,a_3}T_{a_4,a_5,\bar a_3}
T_{\bar a_4,\bar a_2,a_6}T_{\bar a_1, \bar a_4, \bar a_6}$
itself to further reduce these terms. 
Indeed, because the indices of the tensor are
distinguished, colors 1 and 4, 2 and 5, and,  
3 and 6, might be exchangeable. 
A different possible symmetry exchange is among the elements of the couples $(a_l, \bar a_l)$. 
The sum \eqref{sumI} boils down to 64/4= 16 sums.  An algorithm that computes this approach is given in section \ref{sec:algo:better}.

\section{Parallel Computation}
\label{sec:algo}

Our goal is to implement the invariant calculations describe above with algorithms whose performance will scale with increasing parallelism.  The challenge is not only to have efficient implementation of the tensor element operations, but to manage parallel execution with low overhead and good load balance.  First we consider algorithms to perform the invariant calculations.  Then we look at their parallelization. 

\subsection{Naive algorithm}
\label{sec:algo:naive}

The most na\"ive implementation of the contraction described in section \ref{sec:invar:basic} is simply a sum of all the elements of the tensor, multiplied by those of the symplectic matrix.  The algorithm consists of 12 nested loops, and given by Algorithm~\ref{algo:naive}.

\begin{algorithm}
\SetInd{1.5ex}{.1ex}

\KwInput{J: symplectic matrix}
\KwInput{T: tensor}
\KwOutput{Tens: invariant}
Tens = 0 \;
  \For{$a1 \gets0$ \KwTo size \KwBy $1$}{
      \For{$a2 \gets0$ \KwTo size \KwBy $1$}{
        \For{$a3 \gets0$ \KwTo size \KwBy $1$}{
		    A = T[a1][a2][a3]\;
            \For{$b1 \gets0$ \KwTo size \KwBy $1$}{
				TAB = J[a1][b1]\; 
		        \For{$b2 \gets0$ \KwTo size \KwBy $1$}{
                    \For{$b3 \gets0$ \KwTo size \KwBy $1$}{
    					TABB =  TAB * A*T[b1][b2][b3]\;
                        \For{$c1 \gets0$ \KwTo size \KwBy $1$}{
                            \For{$c2 \gets0$ \KwTo size \KwBy $1$}{
									TABC = TABB * J[a2][c2]\;
                                    \For{$c3 \gets0$ \KwTo size \KwBy $1$}{
										TABCC = TABC * T[c1][c2][c3] * J[b3][c3]\; 
                                        \For{$d1 \gets0$ \KwTo size \KwBy $1$}{
											TABCD = TABCC * J[c1][d1]\;
											\For{$d2 \gets0$ \KwTo size \KwBy $1$}{
                                                TABCDD = TABCD * J[b2][d2]\;
                                                \For{$d3 \gets0$ \KwTo size \KwBy $1$}{
                                                    Tens = Tens + TABCDD * T[d1][d2][d3]*J[a3][d3]\;
                                                }
                                            }
                                        }
                                    }
                                }
                            }
                        }
                    }
                }
            }
        }
    }

 \caption{Naive invariant computation.\label{algo:naive}}
\end{algorithm}

\noindent
While there is a high degree of concurrency because of the independence of the operations, the sheer dimensionality of the algorithm ($size^{12}$) results in significant computational scaling challenges as $size$ increases.

\subsection{Exploiting algebraic properties}
\label{sec:algo:better}

Clearly, the complexity of the algorithm given in section \ref{sec:algo:naive} is too high.  Luckily, in the particular case of the symplectic matrix we are considering here (see section \ref{sec:invar:basic}), this complexity can be reduced significantly. The algorithm given by Algorithm~\ref{algo:better} consists of 6 nested loops, instead of 12.  Moreover, each loop is rolling over $size/2$ elements instead of $size$.

    \newbox\tempbox%

\savebox{\tempbox}{
\begin{minipage}[c]{0.45\textwidth}%
\begin{algorithm*}[H]


\SetInd{.6ex}{.05ex}
  \scriptsize
  \hspace*{-3mm}%
 \parbox{\dimexpr\textwidth-\algomargin\relax}{
\KwInput{J: symplectic matrix}
\KwInput{T: tensor}
\KwOutput{Tens: invariant}
Tens=TE=T1=T2=T3=T4=T5=T12 = T13 = T14 = T16 = T23 = T24 = T26 = T123 = T126 = T134 = 0\;
N = size/2\;
\For{$a4 \gets0$ \KwTo N \KwBy $1$}{
    A4 = a4 + N\;
    \For{$a2 \gets0$ \KwTo N \KwBy $1$}{
        A2 = a2 + N\;
        \For{$a6 \gets0$ \KwTo N \KwBy $1$}{
            A6 = a6 + N\;
            W1 = T[a4][a2][a6]\;
            W2 = T[a4][A2][a6]\;
            W3 = T[a4][a2][A6]\;
            W4 = T[A4][A2][a6]\;
            W5 = T[a4][A2][A6]\;
            W6 = T[A4][a2][A6]\;
            W7 = T[A4][A2][A6]\;
            \For{$a1 \gets0$ \KwTo N \KwBy $1$}{
                A1 = a1 + N\;
                \For{$a5 \gets0$ \KwTo N \KwBy $1$}{
                    A5 = a5 + N\;
                    Z1 = T[a1][a5][a6]\;
                    Z2 = T[A1][a5][a6]\;
                    Z6 = T[A1][a5][A6]\;
                    t5 = W3*T[a1][A5][a6]\;
                    tE = W4*T[A1][A5][A6]\;
                    t1 = W3*Z2\;
                    t13 = t1\;
                    t2 = W5*Z1\;
                    t23 = t2\;
                    t3 = W3*Z1\;
                    t4 = W6*Z1\;
                    t12 = W5*Z2\;
                    t14 = W6*Z2\;
                    t134 = t14 \;
                    t16 = W1*Z6\;
                    t24 = W7*Z1\;
                    t26 = W2*T[a1][a5][A6]\;
                    t123 = W5*Z2\;
                    t126 = W2*Z6\;
                    \For{$a3 \gets0$ \KwTo N \KwBy $1$}{
                        A3 = a3 + N\;
                        TE+=tE*T[a1][a2][a3]*T[a4][a5][A3]\;
                        T5+=t5*T[A1][A2][A3]*T[A4][a5][a3]\;
                        X7Y5=T[a1][A2][A3]*T[A4][A5][a3]\;
                        T1 += t1*X7Y5\;
                        T16 += t16*X7Y5\;
                        T2+=t2*T[A1][a2][A3]*T[A4][A5][a3]\;
                        T3+=t3*T[A1][A2][a3]*T[A4][A5][A3]\;
                        T4+=t4*T[A1][A2][A3]*T[a4][A5][a3]\;
                        T12+= t12*T[a1][a2][A3]*T[A4][A5][a3]\;
                        T13+= t13*T[a1][A2][a3]*T[A4][A5][A3]\;
                        T14+= t14*T[a1][A2][A3]*T[a4][A5][a3]\;
                        T23+= t23*T[A1][a2][a3]*T[A4][A5][A3]\;
                        T24+= t24*T[A1][a2][A3]*T[a4][A5][a3]\;
                        T26+= t26*T[A1][a2][A3]*T[A4][A5][a3]\;
                        T123+= t123*T[a1][a2][a3]*T[A4][A5][A3]\;
                        T126+= t126*T[a1][a2][A3]*T[A4][A5][a3]\;
                        T134+= t134*T[a1][A2][a3]*T[a4][A5][A3]\;
                    }
                }
            }
        }
    }
}
Tens = 4*(TE+T12+T13+T14+T16 +T23 +T24 +T26 -(T1 +T2 +T3 +T4 +T5+T123+T126+T134))\;
}
    
\end{algorithm*}%
\end{minipage}}

\begin{algorithm}
\caption{Invariant computation using symmetries of the invariant and algebraic properties of the symplectic matrix.\label{algo:better}}
\clipbox{0pt {\depth} 0pt {\baselineskip}}{\usebox{\tempbox}}\hfill
\raisebox{\depth}{\clipbox{0pt 1ex 0pt {\height}}{\usebox{\tempbox}}}
\end{algorithm}

\subsection{Parallel algorithms}
\label{sec:algo:parall}

\begin{figure}[h!]
    \begin{subfigure}[l]{0.48\textwidth}
      \includegraphics[width=\textwidth]{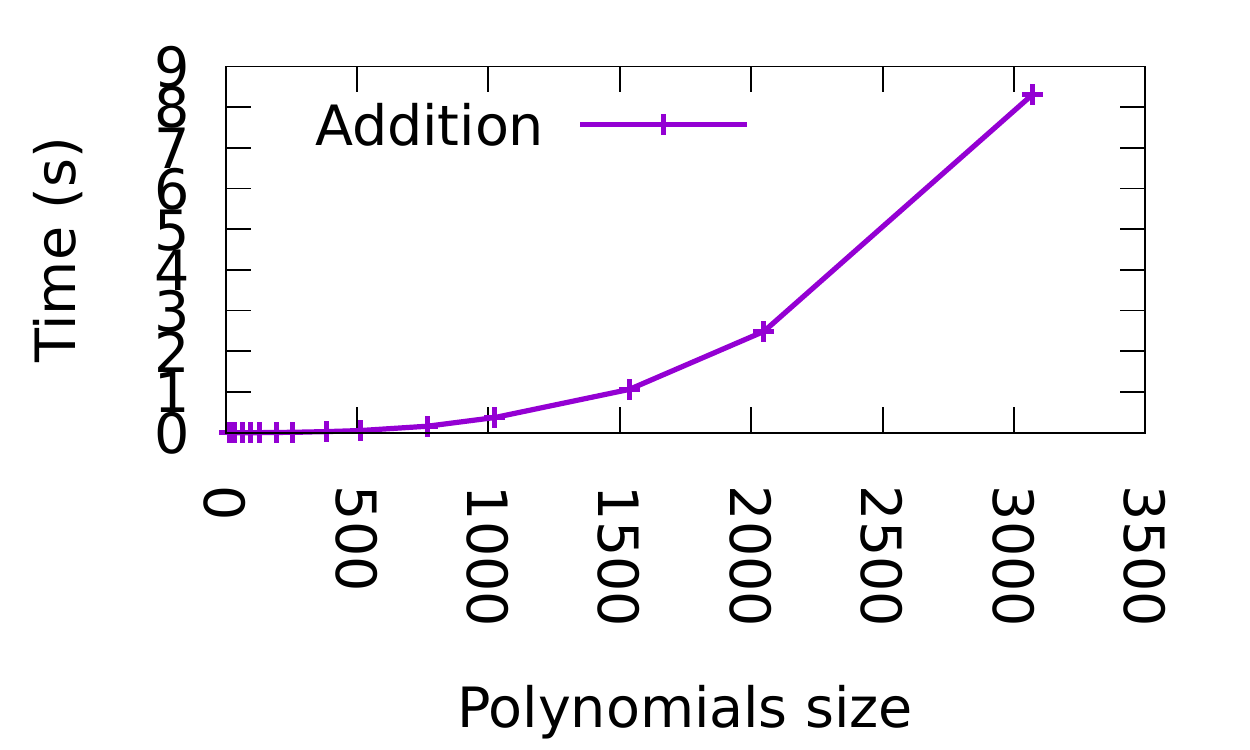}
    \end{subfigure}
    \begin{subfigure}[l]{0.48\textwidth}
      \includegraphics[width=\textwidth]{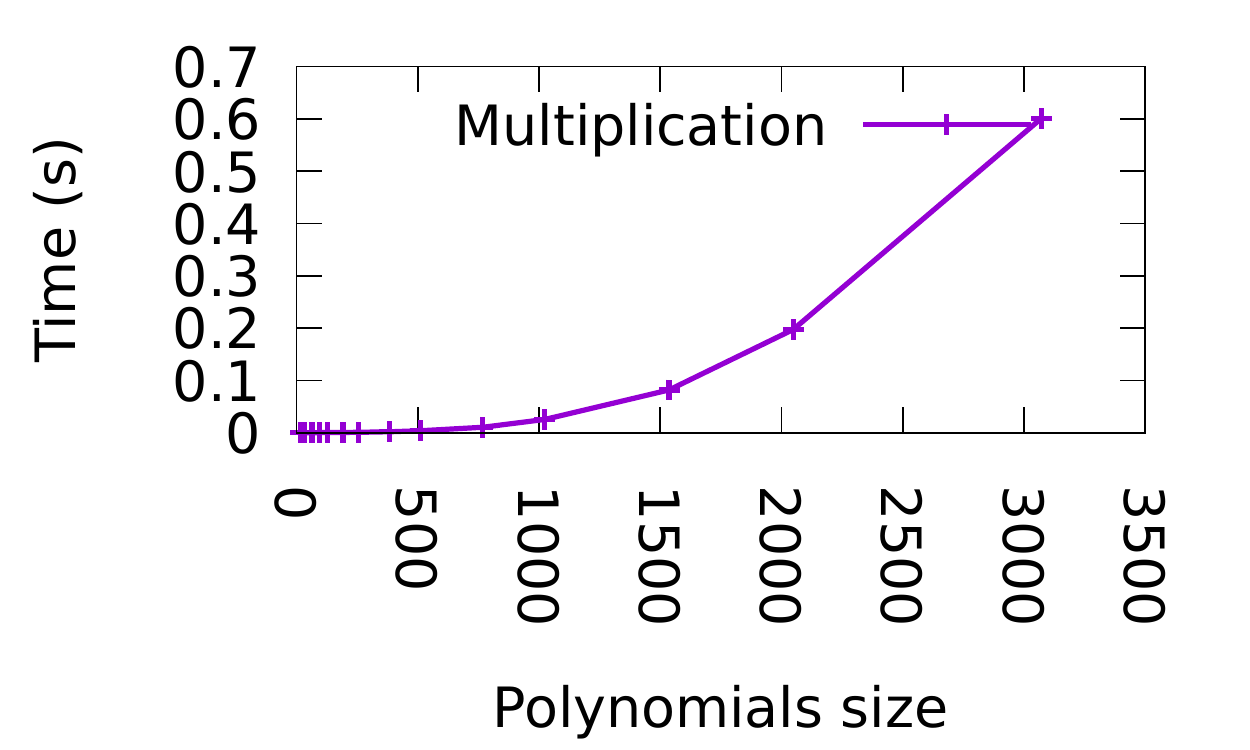}
    \end{subfigure}
    \caption{Cost of basic polynomial operations.\label{fig:poly:cost}}
\end{figure}

We can see that in both algorithms \ref{algo:naive} and \ref{algo:better}, the iterations of the loops are independent from each other, and the final result is a linear combination of the result of each iteration.  Hence, their parallelization might appear straightforward.
However, these algorithms all rely on basic operations (additions, multiplications) on multivariate polynomials.  Such operations have a complexity that depends on the size of the polynomials.  There exist several libraries to perform such operations on symbolic variables, such as GiNaC \cite{ginac}, Piranha \cite{piranha} and its successor Obake \cite{obake}.  We have measured the time taken by the multiplication of a multivariate polynomial by a constant and the addition of two polynomials that have half of their unknowns in common.  The result is presented in Figure \ref{fig:poly:cost}.  Indeed, we can see that the size of the polynomial has a strong impact on the computation time of these basic operations.

As a consequence, we need to be careful about how we design a parallel algorithm to perform the work.  For instance, one strategy could be to statically decompose the iteration of the outer loops into equal parts.  The strategy might be reasonable based on the idea of computing intermediate sums in parallel and then gathering them to produce the final sum.  However, even though subparts of the loops might be of equal lengths with respect to number of iterations to do, they might not have the same execution time.  Although the same number of polynomial operations are performed, these operations might not take the same time, since the size of the polynomials they are manipulating might not be the same, depending on whether terms are canceling each other or not.  Hence, a static domain decomposition scheme (cutting the range of the outer loops in equal parts) might not apply.

Thus, we need to consider schemes whereby the assignment of things to do is more flexible and dynamic.  The question then becomes one of granularity of work assignment to parallel tasks versus the overhead of management and communication with the tasks.  Given the uncertainty of how computationally complex polynomial operations might be, there is also the concern of workload imbalance.  Four parallel algorithms developed in our work are described below.

\paragraph{Master-worker scheme.} The algorithm presented by Algorithm \ref{algo:better} can be implemented in parallel by a master-worker scheme.  It presents the advantage of handling load balancing automatically, since the workers are assigned work to do dynamically.  When a worker requests some work, the master sends it a vector of parameters that contains the indices of the outer loops (hence, the vector's length depends on the granularity) and the worker computes the corresponding inner loops and sends its result to the master.  By linearity, the final result is obtained by addition of all the workers' subresults.  The performance issues that arise in this scheme have to do with the number of works, the overhead of work assignment, and the granularity of the work.

\paragraph{Delegate the addition on a worker.} Adding the intermediate polynomials together can take a significant time, and keep the master busy with computation instead of answering requests from the workers.  A variant of the master-worker scheme consists in only accumulating subresults on the master, and sending two types of workloads to the workers: either a set of parameters, or, when it has accumulated enough subresults, a set of polynomials to add.  This variant reduces the computation workload on the master, but requisitions a worker sometimes during the computation.  The performance issues are similar to the basic scheme, except that more opportunity for work offloading is possible.

\paragraph{Hierarchical master-worker scheme} Depending on the granularity of the computation performed on the workers and the number of workers, the master can be overloaded with communications with the workers.  A hierarchical scheme can be adopted: the worker distributes vectors of parameters to foremen, that cut the corresponding inner loop and distribute them between their workers.  Depending on the granularity, it is expected that the foremen communicate more often with their workers than with the master.  Hence, in addition to reducing the communication bottleneck on the master, this scheme is particularly adapted to hierarchical architectures \cite{CHC09}.

\paragraph{Stateful master-worker} If the bottleneck is on the addition performed by the master to compute the final polynomial, this computation can be distributed.  In a traditional master-worker scheme, the workers are stateless: they send their result to the master and do not keep it in memory.  We can use stateful workers: they do not send their result but instead, they add them.  At the end of the computation, the final result is computed by adding the partial polynomials held by the workers, which can be done using a tree.  Moreover, the addition of the newly computed polynomial to the worker's polynomial can be computed while waiting for the next vector of parameters from the master, hence overlapping communication and computation, and reducing the idle time if the master is overloaded and takes some time to answer.  This algorithm is given by Algorithms \ref{algo:statefulm} (for the master) and \ref{algo:statefulw} (for the worker).

\begin{minipage}[t]{.48\linewidth}
\null 
\begin{algorithm}[H]
\tcc{prepare parameter sets}
queue params\;
\For{$a4 \gets0$ \KwTo N \KwBy $1$}{
    \For{$a2 \gets0$ \KwTo N \KwBy $1$}{
        \For{$a6 \gets0$ \KwTo N \KwBy $1$}{
            params.push\_back( \{ a4, a2, a6 )\} )\;
        }
    }
}
\tcc{distribute them}
\While{ $!parameters.empty()$ }{
    src = recv( request, ANY\_SOURCE )\;
    p = params.pop()\;
    send( src, p, TAG\_WORK )\;
}
\tcc{wait for all the slaves}
\While{ $running()$ }{
    src = recv( request, ANY\_SOURCE )\;
    send( src, 0, TAG\_END )\;
}
\tcc{global sum}
Tens = reduction\_sum()\;
\caption{Master\label{algo:statefulm}}
\end{algorithm}
\end{minipage}%
\begin{minipage}[t]{.48\linewidth}
\null 
\begin{algorithm}[H]
Tens = 0\;
T = 0\;
\While{ true } {
 \tcc{ask for some work}
 send( root, 0, TAG\_REQ )\;
 \tcc{as I wait for a parameter set, add my polynomials}
 req = Irecv( ROOT, ANY\_TAG )\;
 \If{ T != 0  }{
    Tens += T \;
 }
 p, tag = wait( req )\;
 \If{ tag == TAG\_END } {break\;}
 \tcc{compute a polynomial for the parameters I have received}
 T = compute( p )\;
}
\tcc{global sum}
reduction\_sum(Tens)\;
\caption{Stateful worker\label{algo:statefulw}}
\end{algorithm}
\end{minipage}

\subsection{On-the-fly adaptation}
\label{sec:algo:adapt}

All of the parallel schemes described above are reasonable to consider and evaluate.  While there is some support for dynamic work production in their operation, it is not the case that they take into account the cost of doing the work.  Each scheme might be better under difference circumstances.  If the analysis of costs at runtime could allow dynamic selection of which scheme to apply when, it might result in greater performance overall.  There requires performance monitoring of the computation coupled with policy-based online analytics to adapt to changing execution behavior.

\section{Performance Evaluation}
\label{sec:perf}

We have implemented the algorithms described in section \ref{sec:algo} using two symbolic polynomial computation libraries: GiNaC \cite{ginac} and Obake \cite{obake}. We used GiNaC 1.7.6 and its dependency CLN 1.3.4, and Obake commit bbed828 and its dependencies Abseil commit 24713a7, MPPP d56c7502 and MPFR 4.0.2. We have run the performance evaluations on the Grid'5000 platform \cite{grid5000}, using the Parapide cluster in Rennes. It is made of 20 nodes, each of which featuring two Intel Xeon X5570 CPUs (4 cores/CPU), 24 GB of memory and , a 20 Gb InfiniBand NIC and a Giga Ethernet NIC. The operating system deployed on the nodes is a Debian 9.8 with a Linux kernel 4.9.0. All the code was compiled using g++ 8.3.0 with -O3 optimization flag, and OpenMPI 4.0.2. 

\begin{figure}[h!]
\begin{center}
    \includegraphics[width=.8\textwidth]{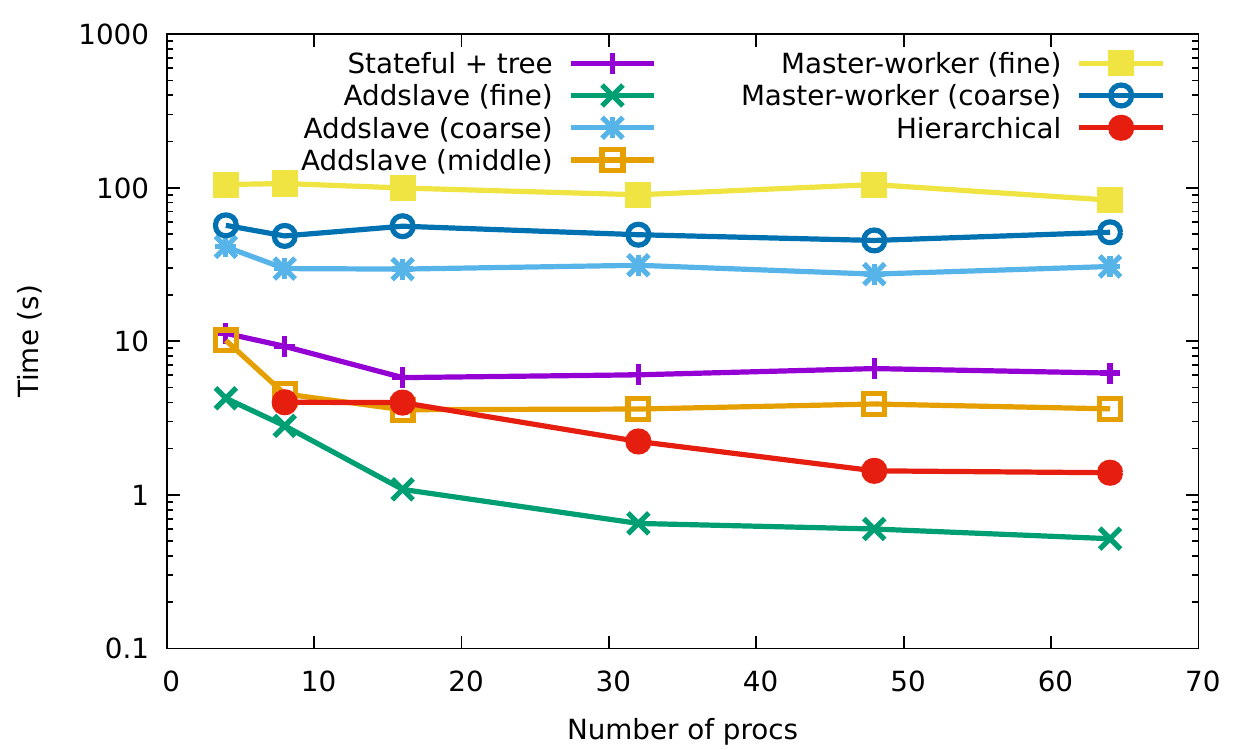}
    \caption{Parallel execution using a tensor of size 8.\label{fig:perf:size8}}
\end{center}
\end{figure}

We compared the execution time of the algorithms described in section \ref{sec:algo:parall} on a small tensor (8 elements in each dimension), on the Parapide cluster, using the Obake library. For readability purpose, since some results were significantly higher than the other ones, we are showing the time in logarithmic scale. We can see that the algorithm that delegates the polynomial additions to a worker, called \emph{addworker} in the remainder of this paper, is significantly faster than the other ones but stops scaling after only 32 processes. On the other hand, the traditional master-worker scheme is slower when using a fine grain. More precise time measurements showed that these computations are limited by the polynomial additions. In the traditional master-worker scheme, the master spends most of its time adding polynomials. When the granularity is finer, it has to perform more additions. The \emph{addworker} takes advantage of the fact that this addition is not performed on the master and therefore, not part of the critical path, but when the number of processes used increases, the final addition becomes the major part of the computation.

Following this observation, we implemented a first adaptation policy, as described in paragraph \ref{sec:algo:adapt}: we introduced some timers on the master in order to determine when the master is taking too much time adding the polynomials compared to the time it waits for workers' results. In other words, we want to detect when the master is not available enough for its workers. We called this algorithm \emph{combined}: it starts as a master-worker and, when it considers it is spending too much time adding the polynomials on the master, it delegates these additions to a worker.

\begin{figure}[h!]
\begin{center}
    \includegraphics[width=.8\textwidth]{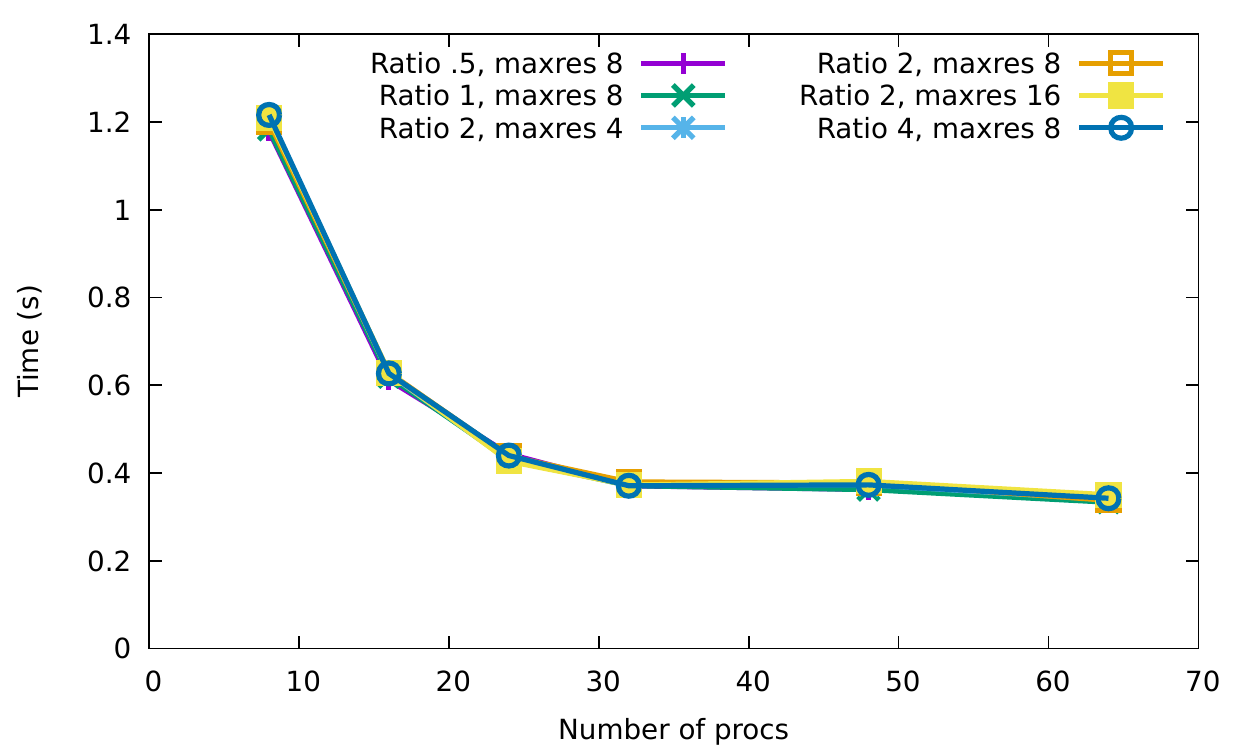}
    \caption{Impact of the policy parameters on the computation, tensor of size 8, on Parapide using Okabe.\label{fig:perf:parameters}}
\end{center}
\end{figure}

The policy that decides when to switch between the tradition master-worker scheme and the \emph{addworker} scheme is parameterized by two constants. The first one is the \emph{ratio} between the computation time when polynomials are added on the master, and the wait time in the reception. The second one is the number of polynomials that are accumulated on the master before sending them to a worker for addition: the \emph{maxresult} constant. We tried to study the impact of these two parameters on a tensor of size 8, using Obake on Parapide. The results are presented on Figure \ref{fig:perf:parameters}. We can see that they do not have any significant impact on the performance.

\begin{figure}[h!]
\begin{center}
    \includegraphics[width=.8\textwidth]{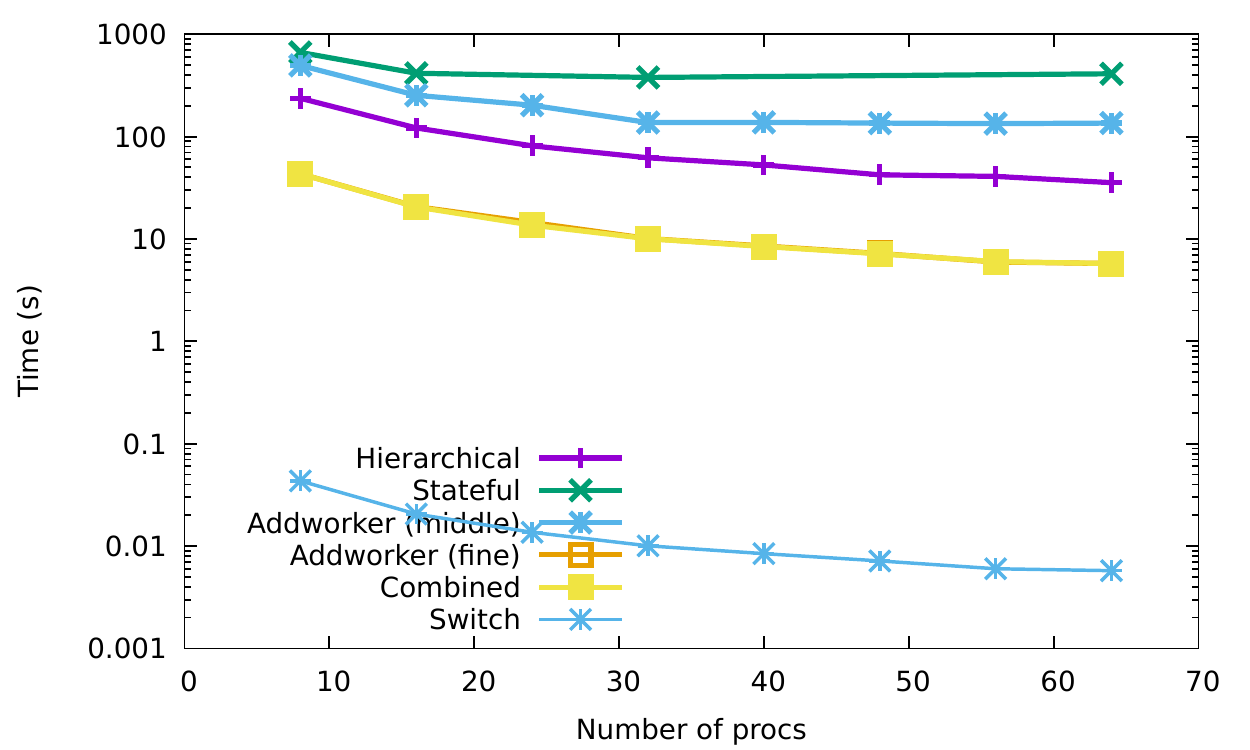}
    \caption{Parallel execution using a tensor of size 12 using Obake.\label{fig:perf:size12:obake}}
\end{center}
\end{figure}

We evaluated it on larger tensors, starting with a tensor of size 12 with Obake (Figure \ref{fig:perf:size12:obake}). We can see that the combined algorithm is slightly faster (1-3\% faster). A closer analysis showed that the algorithm switched from the master-worker to the \emph{addworker} scheme very early in the computation (bottom line, title "Switch"). Hence, since on this problem the polynomials to add are large starting from the beginning of the computation, the algorithm switches early to a scheme which is more efficient in any case.

\begin{figure}[h!]
\begin{center}
    \includegraphics[width=.8\textwidth]{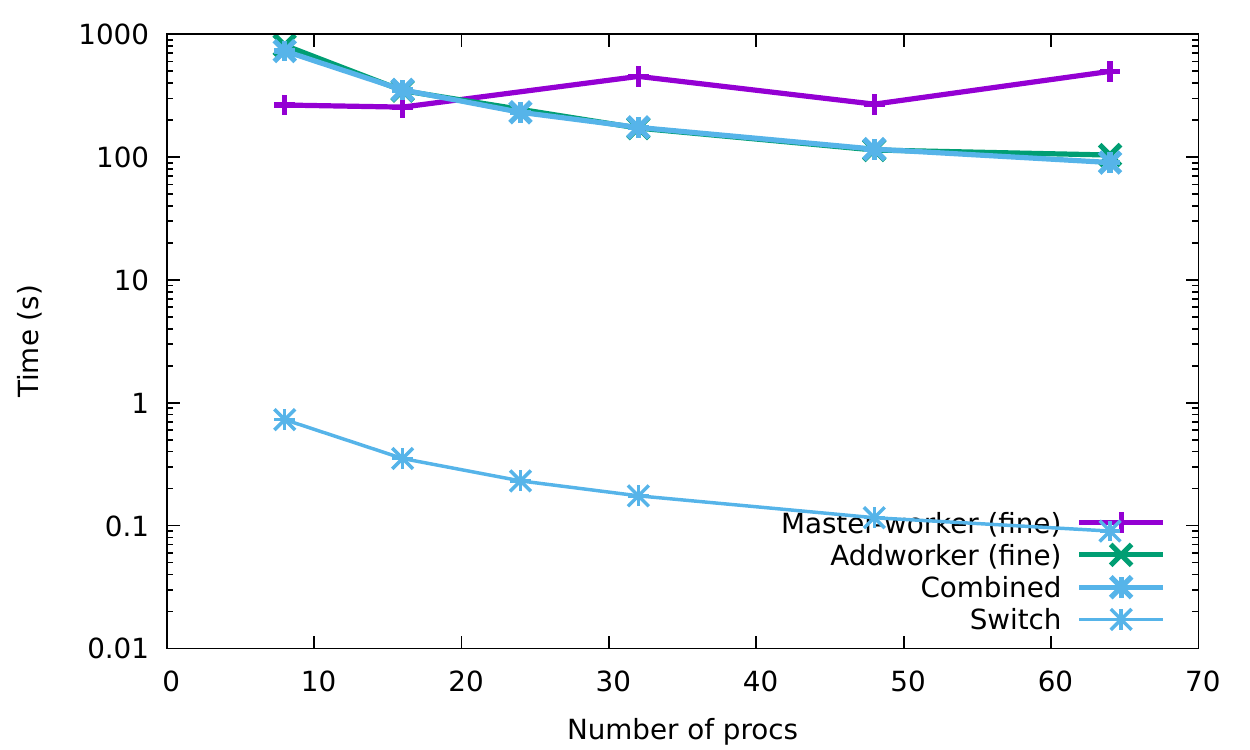}
    \caption{Parallel execution using a tensor of size 16 using Obake.\label{fig:perf:size16:obake}}
\end{center}
\end{figure}

\begin{figure}[h!]
\begin{center}
    \includegraphics[width=.8\textwidth]{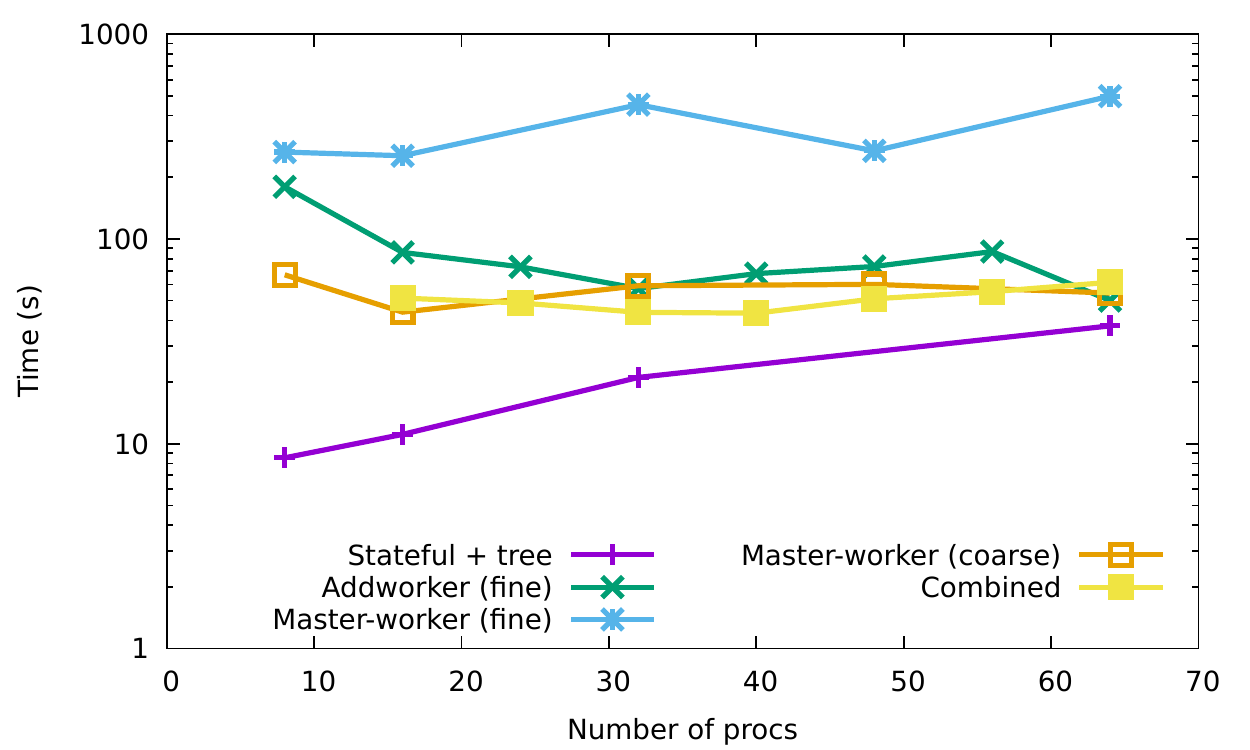}
    \caption{Parallel execution using a tensor of size 16 using GiNaC.\label{fig:perf:size16:ginac}}
\end{center}
\end{figure}

We also measured it on a tensor of size 16 with Obake (Figure \ref{fig:perf:size16:obake}) and with GiNac (Figure \ref{fig:perf:size16:ginac}). GiNaC has an important particularity: the polynomial additions take significantly longer and dominate the computation. Hence, the algorithm that minimizes these additions, \ie the stateful master-worker, performs significantly better but it does not scale because the final addition tree (reduction) dominates the overall execution time. Using Obake, the \emph{addworker} turns out to be less efficient at small scale, and slightly more scalable.

We also implemented a second policy: when the master is overloaded by requests, the computation switches to a hierarchical scheme. The policy change happens after the computation has switched to the \emph{addworker} algorithm. We used timers at two places to determine that the master is a point of congestion: if a worker spends more time waiting for work than computing, and if the master spends more time sending polynomials to add to a worker than waiting for new results. When workers consider they spend too much time waiting for work, they send the next request with a specific tag; if more than half of the workers (over a certain number of iterations) send this tag, the master triggers a change of algorithm. Even when applying a ratio between these timers of the number of processes in the system, we never encountered a situation that triggered this algorithm switch.

Overall, we have seen that:
\begin{itemize}
\item if we increase the number of workers we need to refine the granularity to keep them busy. However, we are reaching a granularity that makes the computation too short compared to the communications;
\item if we increase the size of the problem, we increase the amount of work done by each workers and therefore, we overcome this problem.... but it involves more (expensive) polynomial additions, which increases their share of the overall cost;
\item the polynomial additions become expensive quickly. Switching to the pattern in which the addition is performed by a worker is a good choice most of the times;
\item we have never encountered a case in which we needed to switch to a hierarchical pattern. The reason for that is that the workload on each worker increases faster than the congestion on the master when the size of the problem increases to overcome scaling limitations, and the aforementioned problem of increasing the number of polynomial additions
\item the stateful worker approach is interesting but the final stage that adds all the polynomials, even when done using a tree structure, is expensive, and most of the times its cost is higher than the gain during the computation.
\end{itemize}

\section{Conclusion and Perspective}
\label{sec:conclu}

We began the paper motivating the need for dynamic parallel algorithms that utilize monitoring to guide adaptive control for better performance outcomes.  Indeed, the focus of our work on symplectic invariants in high energy physics is an example of one such application.  There are several features of the problem that translate to requirements for on-the-fly optimization in parallel execution in order to achieve desired performance outcomes.  The parallel algorithms we developed and their implementation with monitoring and adaptive control demonstrate performance improvements over static schemes.  Our experiments show that we can achieve significant results from parallel runs on application cases heretofore unresolved.

Another perspective of this work certainly pertains to physics. Indeed, the fact that the invariant $T^4$ \eqref{T4form} is nontrivial allows us 
to consider it as an interaction in physical models. Tensor models also relate to discrete and random geometry \cite{Guraubook}. The invariant
$T^4$ presented here is associated with a 3d simplex, a tetrahedron that defines the building block of 3D discrete geometries.
We show that, at least for a range of values of $N$, this interaction exists for symplectic tensors and therefore 
opens a new avenue for analysis of tensor models with Sp(N).  Furthermore, and more generally for any symmetry (classical Lie) groups, 
future computational or physical experiments with tensor models will require the data of the invariants itself.  Give it in generic terms of formula
just like $T^4$ \eqref{T4form} will not help.  Hence, extended to other Lie groups, such as the unitary and orthogonal groups, 
the present work already could provide an explicit formula of any polynomial invariant in more than a reasonable time. 
We hope that an efficient calculus of such expressions could contribute to set up those experiments (computational or physical).

We have seen that the parallel performance and scalability we can obtain is always limited by the polynomial operations that need to happen in the critical path. Indeed, a high-performance polynomial calculus library is critical here. Our future work will also consider other forms of parallel execution using multithreading, shared memory, and accelerators, in particular for these polynomial operations.

The source code is available at the following URL: \url{https://depot.lipn.univ-paris13.fr/coti/tensor}.

\bibliographystyle{splncs03}
\bibliography{bibfile}

\end{document}